\documentclass[twocolumn,showpacs,preprintnumbers,amsmath,amssymb,aps,prb]{revtex4}
\usepackage{graphicx}
\begin{document}

\title{Orientational Ordering, Buckling, and Dynamic Transitions for Vortices Interacting with a Periodic 
Quasi-One Dimensional Substrate} 

\author{
Q. Le Thien$^{1,2}$, D. McDermott$^{1,2}$, C. J. Olson Reichhardt$^{1}$,
and C. Reichhardt$^{1}$ 
}
\affiliation{
$^{1}$Theoretical Division,
Los Alamos National Laboratory, Los Alamos, New Mexico 87545 USA\\
$^{2}$Department of Physics, Wabash College, Crawfordsville, Indiana
47933 USA
}

\date{\today}
\begin{abstract}
We examine the statics and dynamics of vortices in the presence of a periodic
quasi-one dimensional substrate, focusing 
on the limit where the vortex lattice constant 
is smaller than the substrate lattice period.  
As a function of the substrate strength and filling factor, within 
the pinned state we observe
a series of order-disorder transitions 
associated with buckling phenomena in which the number 
of vortex rows that fit between neighboring substrate maxima increases. 
These transitions coincide with steps in 
the depinning threshold, jumps in the density of topological defects, 
and changes in the structure factor.
At the buckling transition the vortices are disordered,
while between the buckling transitions
the vortices form a variety of crystalline and partially ordered states.  
In the weak substrate limit, the buckling transitions are
absent and the vortices form an ordered hexagonal lattice that
undergoes changes in its orientation
with respect to the substrate 
as a function of vortex density. 
At intermediate substrate strengths,  
certain ordered states appear that are correlated
with peaks in the depinning force. 
Under an applied drive the system exhibits a
rich variety of distinct dynamical phases, including
plastic flow, a density-modulated moving crystal,
and moving floating solid phases.
We also find a dynamic 
smectic-to-smectic transition in which
the smectic ordering changes from being aligned with the substrate
to being aligned with the external drive.
The different dynamical phases can be characterized
using velocity histograms and the structure factor. 
We discuss how these results are related to recent experiments
on vortex ordering on quasi-one-dimensional
periodic modulated substrates.
Our results should also be relevant for other types of
systems such as ions, colloids, or Wigner 
crystals interacting with periodic quasi-one-dimensional substrates.      
\end{abstract}
\maketitle

\section{Introduction}

Commensurate-incommensurate transitions 
are relevant to a number of condensed matter systems 
that can be effectively described as a lattice of particles interacting
with  
an underlying periodic substrate.  
A commensurate state occurs when certain length
scales of the particle lattice
match the periodicity of the underlying substrate,
such as when the number of particles is equal to the number of 
substrate minima \cite{1,2}. 
Typically when commensurate conditions are met, the system
forms an ordered state free of topological defects, 
while at incommensurate fillings 
there are several possibilities depending on the strength of the substrate. 
If the substrate potential is weak, the 
particles maintain their intrinsic lattice structure
which floats on top of the 
substrate,
while for strong substrates a portion of the particles lock into  
a configuration that is commensurate with the substrate while the remaining
particles form
excitations such as kinks, vacancies, or domain walls.
At intermediate substrate strengths, the lattice ordering can be
preserved
but there can be periodic distortions
or rotations of the particle lattice
with respect to the substrate lattice \cite{3,4,5,6,7,8}.
These different cases are associated with
differing dynamical responses of the particles 
under the application of an external drive \cite{2,7,8,9,10,11}.  
When kinks or domain walls are present,
multi-step depinning
process can occur when
the kinks become mobile at a
lower drive than the commensurate portions of the sample \cite{7,8}.     
Examples of systems that exhibit commensurate-incommensurate phases 
include atoms adsorbed on atomic surfaces \cite{1,3,4}, 
vortices in type-II superconductors
interacting with artificial pinning arrays
\cite{12,13,14,15,16,16a,17,18,19,20}, 
vortex states in Josephson-junction arrays \cite{21,22}, 
superfluid vortices in Bose-Einstein condensates
in the presence of co-rotating optical trap arrays \cite{23,24,25}, 
cold atoms and ions on ordered substrates \cite{26,27,28,29}, 
and colloidal particles on periodic  \cite{6,7,8,30,31,32,33} and
quasi-periodic optical substrates \cite{34,35}.    

In the superconducting vortex system,
commensurability occurs when the number of vortices is an
integer multiple of the number of pinning sites, and
various types of commensurate vortex crystalline states
can occur with different symmetries 
\cite{12,13,16,18}. 
At fillings where there are more vortices than pinning sites,
it is possible to have 
multi-quantized vortices occupy the pinning sites, and
a composite vortex lattice
can form that is comprised of individual or
multiple flux-quanta vortices localized on pinning sites 
coexisting with vortices located in the interstitial
regions between the pinning sites \cite{12,18,19}. 
Ordered commensurate vortex states have been directly
imaged with Lorentz microscopy \cite{13} and other imaging
techniques \cite{36,37},
and the existence of commensuration can also be deduced
from changes in the depinning force needed to move the vortices
when
peaks or steps appear in   
the critical current as a function of vortex density  \cite{12,14,15,16,17,18}. 
It is also possible for ordered vortex structures such as checkerboard states
to form at 
rational fractional commensuration ratios 
of $n/m$ with integer $m$ and $n$, where $n$ is the number of
vortices and $m$ is the number of pinning sites  
\cite{36,37,38,39}. 
Experiments \cite{7} and simulations \cite{8,40} of colloidal
assemblies on optical trap arrays examined the depinning transitions and
subsequent sliding of 
the colloids and show that the
depinning threshold is maximum for one-to-one matching of colloids
and traps, while it
drops at incommensurate fillings due to the presence
of highly mobile kinks, anti-kinks,
and domain walls.

\begin{figure}
\includegraphics[width=3.4in]{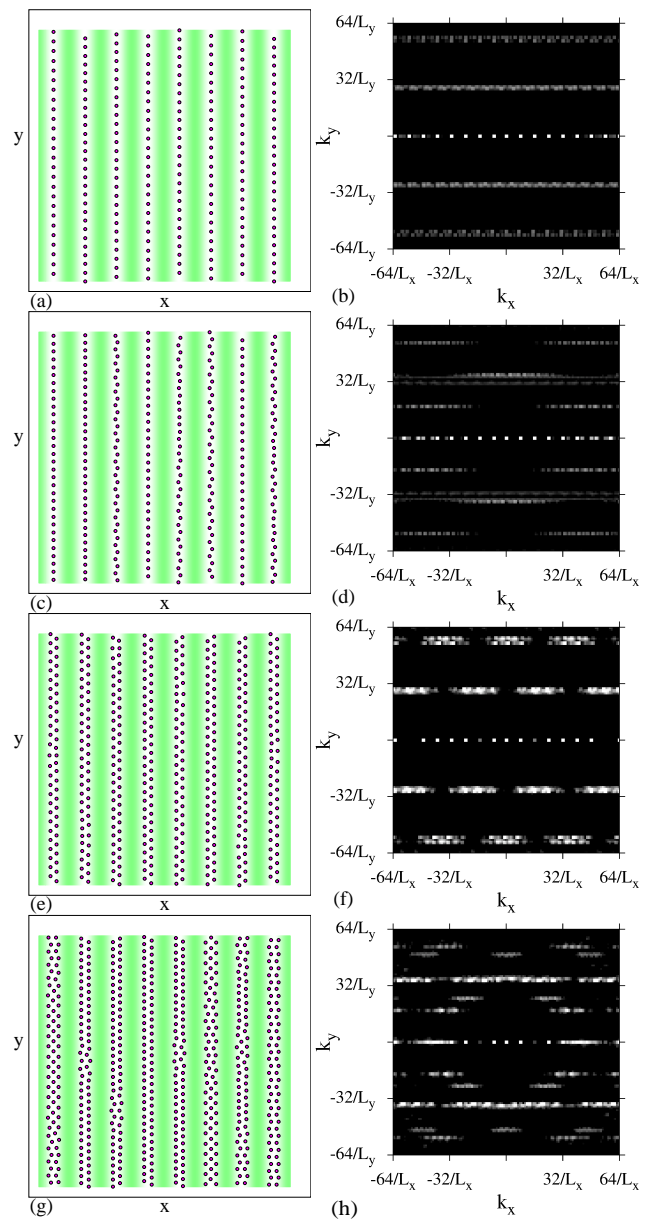}
\caption{The real space images (left column), with the substrate minima
indicated by lighter regions and the vortex positions marked with circles, and
the structure factor $S(k)$ (right column), for a system with a
periodic quasi-one-dimensional substrate with $F_{p} = 1.5$.
(a,b) At $w/a = 1.85$, each substrate minimum contains a single row of
vortices ($r_1$) and the structure factor shows smectic order.
(c,d) $w/a = 2.054$, at the onset of a buckling transition.
(e,f) At $w/a = 2.651$ there is an ordered zig-zag $r_2$ vortex lattice.
(g,h) At $w/a = 3.0$ 
there is a mixture of $r_2$ and $r_3$ lattices.
}
\label{fig:1}
\end{figure}

Commensurate-incommensurate transitions can also occur
for particles
interacting with a periodic quasi-one-dimensional (q1D)
or washboard potential, where 
the particles can slide freely along one direction 
of the substrate but not the other.
An example of this type of system is shown in Fig.~\ref{fig:1} 
for a two-dimensional system of vortices 
interacting with a quasi-one dimensional sinusoidal substrate.
The potential maxima are indicated by the
darker shadings, and the vortices are attracted to the light colored regions.
This type of system has been studied previously for
colloids
interacting with q1D periodic
substrate arrays, where it was shown that various 
melting and structural transitions between hexagonal, smectic,
and disordered colloidal arrangements can occur \cite{41,42,43,44,45,46,47}. 
In general, the colloidal studies focused on the case where the
particle lattice constant $a$ is larger than the substrate
lattice constant $w$. 
Martinoli {\it et al.} investigated vortex pinning 
in 
samples with a 1D
periodic thickness modulation \cite{48,49,50} and observed 
broad commensuration peaks in the depinning threshold that were
argued to be correlated with the formation of ordered vortex
arrangements that could align with the substrate periodicity.  
Other vortex studies in similar samples also
revealed peaks in the critical depinning 
force associated with commensuration effects \cite{51,52},
while studies of vortices interacting with 1D magnetic strips
showed that commensurate conditions were marked by depinning steps
rather than peaks \cite{53}. 
Under an applied dc drive,
depinning transitions occur into a sliding state,
and when an additional ac drive is added to the dc drive, 
a series of Shapiro steps in the voltage-current curves 
appears when the frequency of the oscillatory motion
of the vortex lattice over the periodic substrate locks
with the ac driving frequency \cite{49}.
Similar commensuration effects and Shapiro step phenomena
were also studied for vortices interacting with
periodic washboard potentials or q1D periodic sawtooth substrates \cite{54}.
Vortices interacting with periodic q1D planar defects
have also been studied in layered superconductors  
when the field is aligned parallel to the layer directions.
Here, different vortex lattice structures, smectic states, and oscillations 
in the critical current occur as a function of applied magnetic field 
\cite{55,56,57,58,59,60,61}.

For higher vortex densities in the presence of a q1D substrate
where the vortex lattice constant 
$a$ is {\it smaller} than the substrate lattice constant, $a < w$,
there are several possibilities for how the vortices can order.
In the weak substrate limit, they can form a hexagonal lattice 
containing only small distortions,
while in the strong substrate limit the vortices can
be strongly confined in each potential minimum  to 
form 
1D rows,
so that the overall two-dimensional (2D) vortex structure is anisotropic.
Between these limits, the vortices 
can exhibit buckling transitions by forming
zig-zag patterns within individual potential troughs,
so that for increasing vortex density there could
be a series of transitions at which increasing numbers
of rows of vortices appear in the potential troughs.
Transitions from 1D rows of particles to zig-zag states or multiple rows
have been studied
for particles in single q1D trough potentials 
in the context of vortices \cite{62,63,64},
Wigner crystals \cite{65,66,67}, colloids \cite{69,69,70,71},
q1D dusty plasmas \cite{72,73}, 
ions in q1D traps \cite{74,75,76}, and other systems \cite{77,78}  
where numerous structural transitions, diffusion behavior
and dynamics can occur.
In the case  of a periodic array of channels such as shown in Fig.~\ref{fig:1}, 
much less is known about what buckling transitions would occur and
what the dynamics would be under an applied driving force. 
Recently  Guillam{\' o}n {\it et al.} studied vortex lattices in samples with 
a periodic q1D array of grooves. 
As a function of the commensuration ratio $p = w/a$,
they found that for  $p < 6$, the vortex lattice remains
triangular but undergoes a series of 
transitions
that are marked by rotations of the angle $\theta$ made by
the vortex lattice with respect to the substrate
symmetry direction \cite{79}. 
They also observed that at much higher fields, the system transitions
into a disordered state with large vortex density fluctuations.
Open questions include what happens to these reorientation transitions
as the substrate strength is increased, and what 
the vortex dynamics are when a driving force is applied.
Dynamical phases and structural transitions between different kinds
of nonequilibrium vortex flow states have been extensively studied for 
driven vortex systems interacting with random \cite{80,81,82,83,84,85}
and 2D periodic pinning arrays \cite{86,86a,87};
however, there is very little work examining the dynamic vortex phases
for vortices moving over q1D periodic substrates.
It is not known whether the vortices would
undergo dynamical structural transitions or exhibit the same types
of dynamic phases found for vortices driven over random 
disorder, such as a disordered plastic flow state that transitions to
a moving smectic or anisotropic crystal as a function of increasing drive.   

In this work we consider ordering and dynamics of vortices
interacting with a periodic q1D sinusoidal potential for 
fillings $0 < w/a < 5.5$. In 
the strong substrate regime, the system undergoes a series of
structural transitions 
that are related to the number of rows $r_n$ of vortices
that fit within each substrate trough.
These transitions include
transformations from 1D vortex rows
to zig-zag patterns that gradually increase $r_n$.
The vortex structure contains numerous dislocations
at the buckling transitions and is ordered between the buckling transitions.  
For strong substrates, the onset of the
buckling and ordered phases produces a series of steps in the critical depinning
threshold as a function of vortex density,
while for weaker substrate strengths, some of the states
in which the vortices order 
produce peaks in the critical depinning force.
For the weakest substrates,
the vortices form a triangular lattice
that undergoes rotations with respect to the underlying
substrate symmetry direction as a function of applied magnetic
field, similar to the behavior observed by Guillamon {\it et al.} \cite{79}.  
Under an applied drive we observe  
plastic flow states, moving density-modulated crystals, and dynamic
floating solids.  
For certain fillings we also find smectic-to-smectic transitions where the
two smectic states have
different orientations.
These different flowing phases produce distinct features in
the velocity histograms  
and the structure factor.   
The commensurability ratio $w/a$ also strongly affects the driving force at
which the transition to a moving floating solid occurs.       
Our results should be general to other types of systems that can be
represented as a collection of repulsive particles interacting with a periodic 
q1D substrate, such as colloids on optical line traps, ions in coupled traps,
and Wigner crystals on corrugated substrates.  

\section{Simulation} 
We model a two-dimensional system of 
vortices interacting with a periodic q1D potential
with period $w$, where there are 
periodic boundary conditions in the $x$ and $y$-directions. The vortices
are modeled as point particles and the dynamics of an individual vortex
$i$ obeys the following equation of motion:
\begin{equation}
\eta \frac{ d {\bf R}_{i}}{dt}  = {\bf F}^{i}_{vv} + {\bf F}^{s}_{i} + {\bf F}^{i}_{d} + {\bf F}^{i}_{T}.
\end{equation} 
Here
$\eta$ is the damping constant which we set equal to unity.
The vortex-vortex forces
${\bf F}^{i}_{vv} = \sum^{N_{v}}_{j=1}F_{0}K_{1}(R_{ij}/\lambda){{\bf \hat R}_{ij}}$,
where $F_{0} = \phi_{0}^{2}/2\pi\mu_{0}\lambda^3$,
$\phi_{0}$ is the elementary flux quantum,
$\mu$ is the permittivity, $K_{1}$ is the modified Bessel function,
${\bf R}_{i}$ is the location of
vortex $i$, $R_{ij} = |{\bf R}_{i} - {\bf R}_{j}|$,
${\bf \hat R}_{ij}=({\bf R}_i-{\bf R}_j)/R_{ij}$,
and $\lambda$ is the penetration depth.  
The vortices have repulsive interactions and form a triangular lattice
in the absence of a substrate.
The vortex interaction with the substrate is given by
${\bf F}^{s}_i  = -\nabla V(x_i) {\bf \hat x}$
where the substrate has the sinusoidal form  
\begin{equation}
V(x)  = V_{0}\sin(2\pi x/w).
\end{equation}
We define the pinning strength of the substrate to be
$F_{p} = 2\pi V_{0}/w$. 
The dc driving force ${\bf F}^{i}_{d}$ arises from the Lorentz force
induced by a current applied along the easy direction ($y$-axis) of the
substrate
which produces a perpendicular force on the vortices and causes them
to move in the 
$x$-direction in our system.
We measure the vortex velocity
$\langle V_x\rangle$ along the driving direction
as we increase the external drive in increments of
$\delta F_{d}$, and average 
the vortex velocities over a fixed time in order to avoid any
transient effects.
The thermal forces ${\bf F}_{T}$ are modeled as random Langevin kicks with
the properties $\langle {\bf F}_{T}\rangle = 0$
and
$\langle {\bf F}^{i}_{T}(t){\bf F}_T^j(t^{\prime})\rangle = 2\eta k_{B}T
\delta_{ij}\delta(t - t^{\prime})$,
where $k_{B}$ is the Boltzmann constant.
The initial vortex positions
are obtained by annealing from a high temperature state and cooling
down to $T = 0$.
The dc drive is applied only after the annealing procedure is completed.
We consider a range of 
vortex densities, which we report
in terms of the ratio $w/a$ of the periodicity of the substrate
to the vortex lattice constant that would appear in the absence of a substrate.
We denote a state containing $n$ rows of vortices in each potential
minimum as $r_n$.

\section{Pinned Phases}

In Fig.~\ref{fig:1}(a,c,e,g), we plot the real space locations of the
vortices on the potential substrate
after annealing for a system with $F_{p} = 1.5$ at fillings of
$w/a = 1.85$,  2.054, 2.651, and $3.0$,
while in Fig.~\ref{fig:1}(b,d,f,h) we show the corresponding structure
factors $S(k)$.
At $w/a = 1.58$ in Fig.~\ref{fig:1}(a), the vortices form single 1D
rows in each potential minimum, corresponding to an $r_1$ state,
and the overall vortex structure is highly 
anisotropic with lattice constants $a_x = 4.5$ in the $x-$direction and 
$a_{y} = 1.31$ in the $y-$direction.
Additionally, each potential trough captures a slightly different number of
vortices, introducing disorder in the alignment of
rows in adjacent minima, and leaving the system with periodic ordering
only along the $x$-direction.
The corresponding structure factor in Fig.~\ref{fig:1}(b)
exhibits a series of spots at $k_{y} = 0$ and finite  $k_{x}$,
indicative of the 1D ordering associated with a smectic phase.
As the magnetic field increases,
the vortex ordering must become
increasingly anisotropic in order to maintain
single rows of particles in each minimum.  This is energetically
unfavorable, so instead a
transition occurs to a zig-zag or buckled state
in which there are two partial rows of vortices in every substrate minimum. 
Figure~\ref{fig:1}(c) illustrates the real space
vortex positions for $w/a = 2.054$ at the beginning of the zig-zag
transition, where some of the troughs contain a buckled vortex pattern.
In the corresponding $S(k)$ plot in Fig.~\ref{fig:1}(d),
the smectic ordering develops additional features at large $k$ associated with
the shorter range structure that arises on the length scale associated
with the zig-zag pattern.
As the magnetic field is further increased, 
the zig-zag pattern appears in all the substrate minima and the system
forms an ordered anisotropic 2D $r_2$ lattice 
as shown in Fig.~\ref{fig:1}(e) for
$w/a = 2.651$,
where there are two rows of vortices in each potential
minimum that form a zig-zag structure 
which is aligned with zig-zag structures in neighboring minima.
The corresponding $S(k)$ in Fig.~\ref{fig:1}(f)
has a series of 
peaks at small and large $k$ indicating the
presence of a more ordered vortex structure.
For higher fields, the zig-zag lattice becomes
increasingly anisotropic until another buckling transition
occurs to produce $r_3$ with three vortex rows per substrate minimum.
Figure~\ref{fig:1}(g) shows the transition point at $w/a = 3.0$
where certain potential troughs contain three vortex rows
while others contain two vortex rows or mixtures
of two and three vortex rows.
In Fig.~\ref{fig:1}(h),
$S(k)$ for this case
shows that the system is considerably more disordered than
at the commensurate case illustrated in Fig.~\ref{fig:1}(e,f). 

\begin{figure}
\includegraphics[width=3.5in]{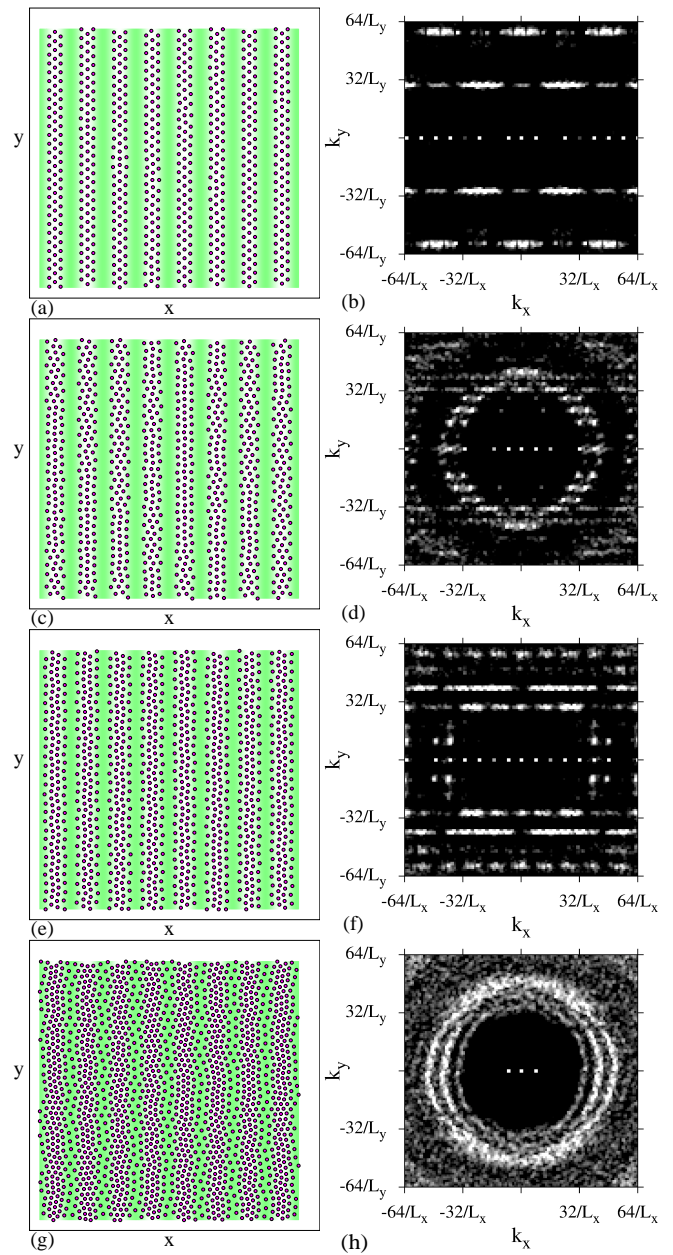}
\caption{The continuation of the  real space images
(left column) and $S(k)$ (right column) from the system
in Fig.~\ref{fig:1} with $F_p=1.5$.
(a,b) At $w/a = 3.3$, there is an ordered structure with three vortex
rows per potential minimum ($r_3$).
(c,d) At $w/a = 3.644$, there is a partially disordered state with
roughly three vortex rows per potential minimum.
(e,f) At $w/a = 4.1455$ there is a partially ordered state
with four vortex rows per minimum ($r_4$).
(g,h) The disordered state at $w/a = 5.15$ showing ring structures in $S(k)$. 
}
\label{fig:2}
\end{figure}

In Fig.~\ref{fig:2} we show the continuation of the evolution of the
vortex lattice from Fig.~\ref{fig:1} in both real space and $k$-space.
At $w/a=3.3$ in
Fig.~\ref{fig:2}(a),
there is an ordered $r_3$ structure with three vortex rows in each potential
minimum, producing
the ordered $S(k)$ shown in Fig.~\ref{fig:2}(b).
As the vortex density is further increased,
the row structure disorders as
shown in 
Fig.~\ref{fig:2}(c) for $w/a = 3.644$,
corresponding to a ring like structure in $S(k)$
as indicated in Fig.~\ref{fig:2}(d).
There are still peaks along the $k_{y} = 0.0$
line due to the anisotropy induced by the substrate. 
For this value of $F_p$, further increasing the vortex density does
not produce a more ordered configuration; however,
certain partially ordered structures can occur
as illustrated in Fig.~\ref{fig:2}(e) for
$w/a = 4.1455$, where there are four vortex rows per trough ($r_4$)
with mixed peaks and smearing in the corresponding structure factor
shown in Fig.~\ref{fig:2}(f).
At higher fields, the vortex structures become disordered 
as shown in Fig.~\ref{fig:2}(g) at $w/a = 5.15$,
where $S(k)$ in Fig.~\ref{fig:2}(h) has pronounced ring structures.
There are still two peaks at $k_y=0$ and finite $k_x$
due to the smectic ordering imposed by the q1D substrate.

\begin{figure}
\includegraphics[width=3.5in]{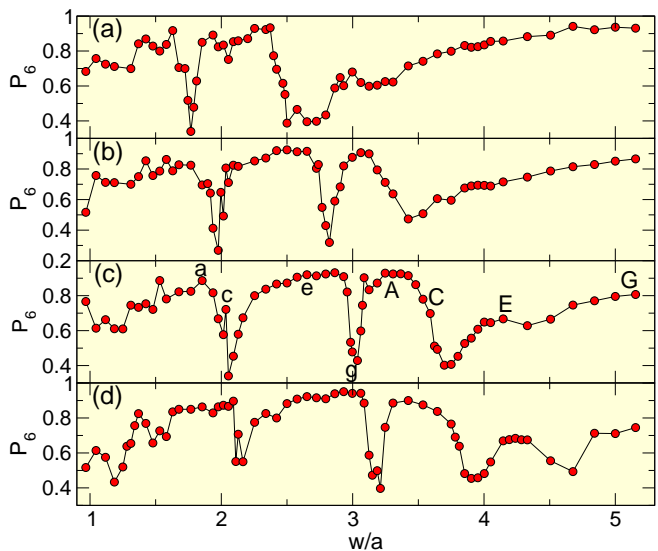}
\caption{ The fraction of six-fold coordinated vortices $P_{6}$  vs $w/a$
for (a) $F_{p} = 0.5$, (b) $F_{p} = 1.0$,
(c) $F_{p} = 1.5$ and (d) $F_{p} = 2.0$. 
In panel (c), the labels a, c, e, g indicate the values of $w/a$ at which
the images in Fig.~\ref{fig:1} were obtained, while the labels
A, C, E, and G indicate the values of $w/a$ at which the images in
Fig.~\ref{fig:2} were obtained.
The dips in $P_{6}$ coincide with
transitions in the number of vortex rows contained
within each potential minimum.   
}
\label{fig:3}
\end{figure}

We can also characterize the system using
the fraction of six-fold coordinated vortices
$P_{6}=N^{-1}\sum_{i=1}^N\delta(6-z_i)$, where $z_i$ is the coordination
number of vortex $i$ obtained from a Voronoi construction.
In general, we find that $P_6$ drops
at the buckling transitions due to the formation
of dislocations that are associated with the splitting of
a single row of vortices into two rows,
creating a kink at the intersection of the two rows.
In Fig.~\ref{fig:3}(a) we plot $P_{6}$ versus $w/a$ for a system 
with $F_{p} = 0.5$.
Over the range
$1.0 < w/a < 1.7$, each pinning trough contains an $r_1$ state,
while the dip in $P_{6}$ at $w/a = 1.77$ corresponds 
to the middle of the buckling transition when there is roughly  
a 50:50 mixture of $r_1$ and $r_2$.
For $ 1.85 < w/a < 2.35$,
the system forms an
ordered $r_2$ state similar to that shown in
Fig.~\ref{fig:1}(e), but less anisotropic since the weaker substrate
compresses the zig-zag structure less and permits it to be wider.
Near $w/a = 2.4$, there is another buckling transition
from $r_2$ to $r_3$
and the system
forms a disordered state similar to that
shown in Fig.~\ref{fig:1}(g).
As $w/a$ is further increased, there is a partially ordered state
near $w/a = 3$ that is similar to the state in
Fig.~\ref{fig:2}(a); however, due to the weaker substrate strength
a fully ordered $r_3$
state does not form.
For $w/a > 3.2$ the system adopts a polycrystalline configuration that
becomes more ordered at high vortex densities.
In Fig.~\ref{fig:3}(b), we show that a similar set of
features associated with buckling transitions occurs
for a stronger substrate with $F_{p} = 1.0$;
however, in this case 
the transition from $r_2$ to $r_3$ 
is sharper
and a fully ordered three row state appears
near $w/a = 3.0$.

In Fig.~\ref{fig:3}(c) we plot $P_6$ versus $w/a$ for
samples with $F_{p} = 1.5$, the same pinning strength at which the images
in Figs.~\ref{fig:1} and \ref{fig:2} were obtained.
Here the dips in $P_6$ associated with the $r_1$ to $r_2$, $r_2$ to $r_3$,
and $r_3$ to $r_4$ transitions
are sharper.
We also observe the development of a small dip near $w/a = 4.4$ corresponding
to a partial transition from $r_4$ to $r_5$.
The values of $w/a$ at which row transitions occur shift upward 
with increasing $F_{p}$. 
For example, the $r_1$ to $r_2$ transition occurs at
$w/a=1.768$ for $F_p=0.5$ but at $w/a=2.05$ for $F_p=1.5$, since the
higher $F_p$ stabilizes the $r_1$ state up to higher
lattice constant anisotropies.
Figure~\ref{fig:3}(d) shows $P_6$ versus $w/a$ for
samples with $F_{p} = 2.0$. Here the dip in $P_6$
at the $r_1$ to $r_2$ transition broadens,
while a pronounced jump emerges at $w/a = 4.7$ corresponding
to the $r_4$ to $r_5$ transition.
We expect that for higher values of $F_{p}$, additional dips in $P_6$
for transitions from $r_n$ to $r_{n+1}$ states for $n\geq 5$ will appear
at $w/a$ values higher than those we consider here.

It is difficult to determine if the buckling transitions are first or second
order in nature.
For particles 
in an isolated trough, the  transition
from a single row to a zig-zag pattern is second order,
and there have been several studies in cold ion systems of
quenches through this transition
in which the density of kinks was calculated for different quench rates
and compared to predictions from nonequilibrium
physics on quenches through continuous phase transitions \cite{74,75,76}.
We expect that the buckling transitions we observe are second order;
however, it may be possible that the additional coupling to particles
in neighboring potential minima
could
change the nature of the transition, and
we have observed a coexistence of 
chain states which is suggestive of phase separation.
For vortex systems it could be difficult to change the substrate strength 
as a function of time,  but for colloidal systems
it is possible to create q1D periodic optical substrates of adjustable
depth and use them to study
time dependent transitions by counting the number of kinks that form
as a function of the rate at which the substrate
strength is changed.

\begin{figure}
\includegraphics[width=3.5in]{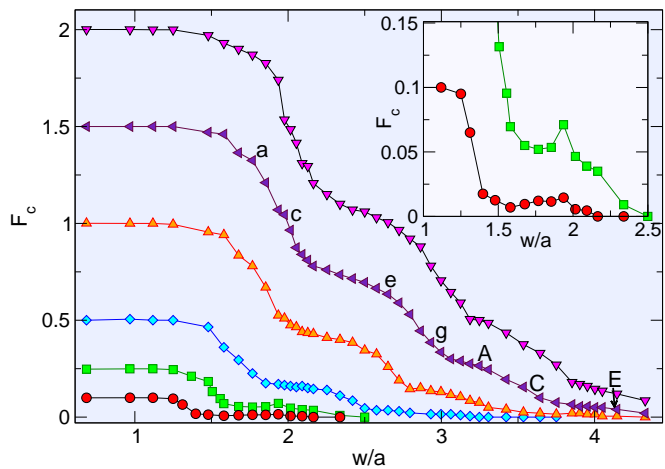}
\caption{ The depinning force $F_{c}$ vs $w/a$ for
$F_{p} = 0.1$ (red circles),
0.25 (green squares), 0.5 (blue diamonds), 1.0 (orange up triangles),
1.5 (purple left triangles),
and 2.0 (pink down triangles)  
showing that for $F_{p} > 0.25$ the buckling transitions
correspond to step features in $F_c$.
The labels a, c, e, g indicate the values of $w/a$ at which the
images in Fig.~\ref{fig:1} were obtained, while the labels
A, C, and E indicate the values of $w/a$ at which the images in
Fig.~\ref{fig:2} were obtained.
Inset: a highlight of the main panel illustrates that
for weaker pinning, peaks in $F_{c}$ occur, as shown for
$F_p=0.1$ (red circles) and 0.25 (green squares).
The peak is associated with the
formation of an ordered zig-zag lattice similar to that
shown in Fig.~\ref{fig:1}(e).  
}
\label{fig:4}
\end{figure} 

\begin{figure}
\includegraphics[width=3.5in]{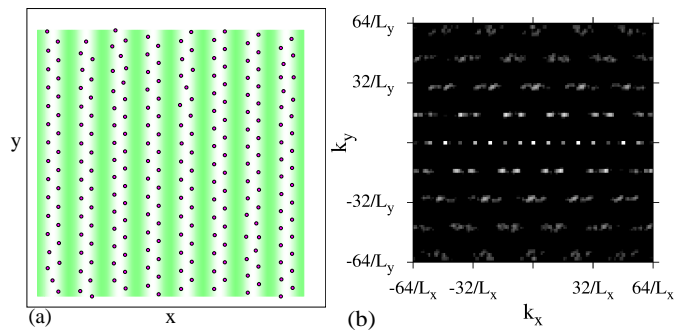}
\caption{(a) Real space image of
the vortex configuration at the peak in
$F_{c}$ at
$F_{p}= 0.25$ and $w/a = 1.94$  for the system shown in the 
inset of Fig.~\ref{fig:4} where an ordered zig-zag structure occurs.
(b) The corresponding $S(k)$ contains various peaks
reflecting the ordered nature
of the state.    
}
\label{fig:5}
\end{figure}

In Fig.~\ref{fig:4} we plot the depinning force $F_{c}$
versus $w/a$ for $F_{p} = 0.1$, 0.25, 0.5, 1.0, 1.5, and $2.0$ to show
that the buckling transitions are associated with
changes in the slope of the depinning force,
which decreases with increasing $w/a$ in a series of steps.
The first drop in $F_c$ near $w/a = 2.0$ corresponds to the
$r_1$ to $r_2$ transition.
In the $r_1$ state, the particle-particle interactions
roughly cancel in the $x$-direction, so the depinning force 
is approximately equal to $F_{p}$,
while close to the buckling transition
the vortices on the right side of a zig-zag
experience an additional repulsive force in the driving direction
from the vortices on the
left side of the zig-zag, decreasing the driving force needed to
depin the vortices.
To estimate the magnitude of this reduction in $F_{c}$, 
we note that the average $x$-direction spacing between vortices
in a given zig-zag
is approximately  
$d_x=2.0$.
For a zig-zag with a 30$^\circ$ angle between the two closest neighbors
on the other side of the chain from each vortex,
the vortex-vortex interaction force of $K_{1}(2)$ 
produces an additional repulsive force of
$f_r=0.55$, giving a value of $F_c=F_p-f_r$ that is close to the
value of $F_c=0.78$ observed after the
$r_1$ to $r_2$ step for the $F_p=1.5$ system.
Similar arguments can be made for the magnitudes of the drops in $F_c$
at the higher order transitions as well.
At $w/a = 1.85$ in the $F_p=1.5$ sample,
$F_c$ already begins to drop below $F_p$
even though the pinned configuration
shown in Fig.~\ref{fig:1}(a) is an $r_1$ state.
This occurs because
in this range of $w/a$, application of a finite $F_d<F_c$ induces
a slight buckling of the vortices,
while for $w/a < 1.5$ the
$r_1$ rows remain in a 1D pinned state up to $F_{d} = F_{c}$.
The inset of Fig.~\ref{fig:4} shows a blowup of
$F_{c}$ versus $w/a$ for the weaker pinning cases $F_{p} = 0.25$ and $F_p=0.1$.
At $w/a=1.94$ there is a peak in $F_c$ for the $F_p=0.25$ sample 
coinciding with
the formation of the long range ordered zig-zag state shown in
Fig.~\ref{fig:5}(a).
The corresponding
$S(k)$ in Fig.~\ref{fig:5}(b) contains a series of peaks that reflect
the ordered nature of the state, which resembles
the zig-zag state in Fig.~\ref{fig:1}(e,f) except
that the system is more ordered
and the zig-zag structure is wider.
For $F_{p} = 0.1$ the zig-zag state transitions into a
hexagonal lattice and the peak in $F_c$ begins to disappear.
Some experiments
examining vortices in q1D periodic pinning structures 
show that peaks in the critical current
occur at certain fillings \cite{48,50,51,52} in regimes where the
pinning is weak, whereas
other experiments performed in the strong pinning limit reveal
more step-like features in the critical current.  This suggests that
the experiments in the strong pinning limit
are producing buckling transitions \cite{53}. 

\begin{figure}
\includegraphics[width=3.5in]{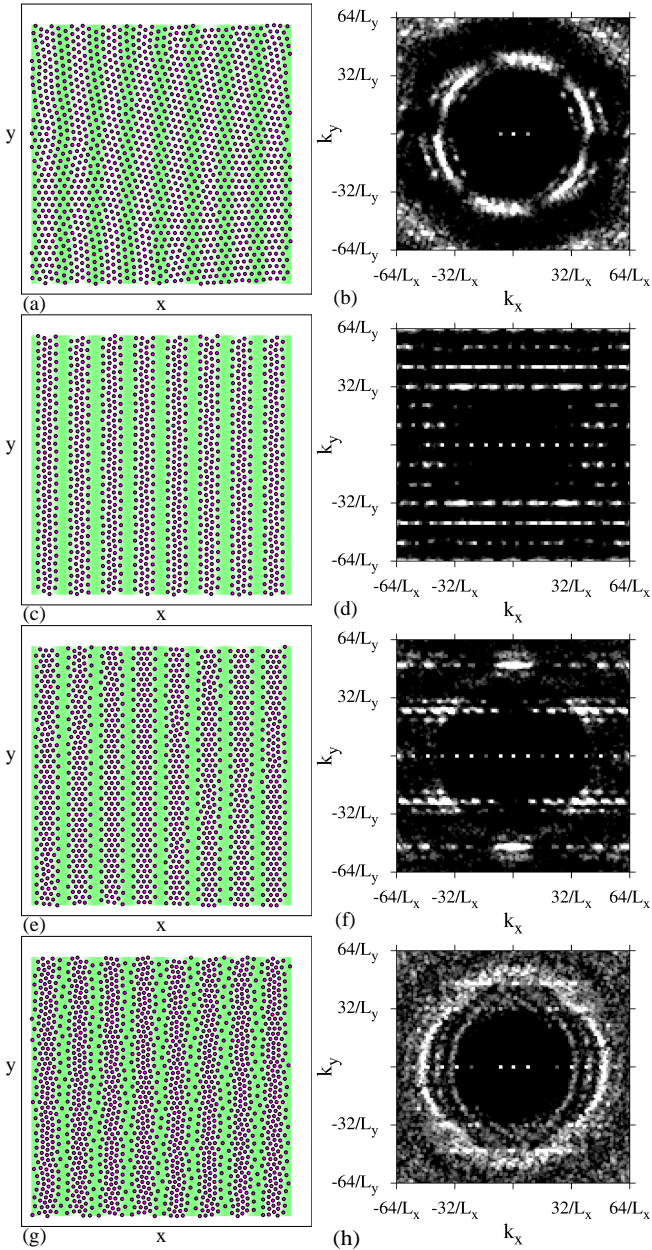}
\caption{Real space images (left column), with the substrate minima
indicated by lighter regions and the vortex positions marked with circles,
and $S(k)$ (right column).  
(a,b) At $F_{p} = 0.5$ and $w/a = 4.825$ there is a polycrystalline structure. 
(c,d) At $F_{p} = 2.0$ and $w/a = 4.33$ there is a partially ordered
$r_4$ structure.
(e,f) At $F_{p}= 2.0$ and $w/a = 4.67$, an ordered structure appears.
(g,h) At $F_{p} = 2.0$ and $w/a = 5.15$ the structure is disordered.
}
\label{fig:6}
\end{figure}

In Fig.~\ref{fig:6} we plot representative 
real space images with the matching $S(k)$ for some other substrate
strengths to highlight other types of ordering we observe.
Figure~\ref{fig:6}(a) shows the real space ordering of
the vortices at $F_{p} = 0.5$ and $w/a = 4.825$, where the vortex
lattice is polycrystalline and contains
regions of triangular ordering
with different orientations.
The corresponding structure factor in Fig.~\ref{fig:6}(b)
has ring features 
with some remnant of the smectic ordering
appearing at smaller values of $k$.  
In Fig.~\ref{fig:6}(c), 
at $F_{p} = 2.0$ and $w/a = 4.33$ an $r_4$ state appears, while
$S(k)$ in
Fig.~\ref{fig:6}(d) has smectic ordering features along with
additional crystalline ordering signatures
due to the ordered arrangement of the particles within the troughs.
At $F_{p} = 2.0$ and $w/a = 4.67$ in
Fig.~\ref{fig:6}(e),
a new type of ordered structure appears in which
the vortices can pack more closely 
by forming alternate regions of $r_3$ and $r_4$ states,
producing a considerable amount of  
triangular ordering
as
seen in the plot of $S(k)$ in Fig.~\ref{fig:6}(f),
where there are sixfold peaks at large $k$ and smectic peaks at smaller $k$.
In Figs.~\ref{fig:6}(g,h), for
$F_{p} = 2.0$ 
and $w/a = 5.15$, a more disordered structure appears, with some regions
of the sample containing $r_4$ or $r_5$ states.

\section{Lattice Rotations for Weak Substrates}

\begin{figure}
\includegraphics[width=3.5in]{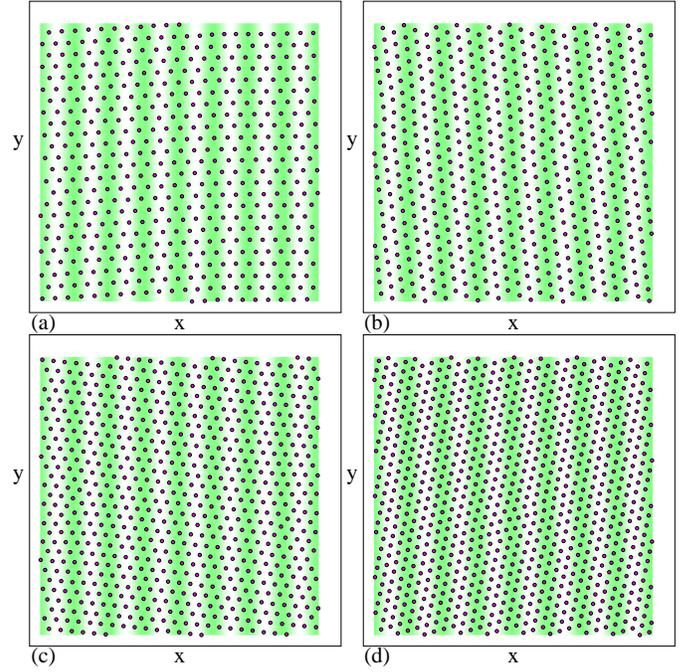}
\caption{Real space images of the vortices in a sample
with $F_{p} = 0.02$ showing that the hexagonal vortex lattice
can adopt various orientations $\theta$ with respect to the substrate.
(a) At $w/a = 2.8$, $\theta = 27.6^{\circ}$.
(b) At $w/a = 3.272$, $\theta = 24^{\circ}$.
(c) At $w/a = 3.53$, $\theta  = 2.9^{\circ}$.
(d) At $w/a = 3.75$, $\theta = 13.2^{\circ}$.  
}
\label{fig:7}
\end{figure}

We have also studied systems with a pinning
strength of $F_{p} = 0.02$.
Here, for $w/a > 1.77$ the vortices 
form a triangular lattice and the features
associated with the buckling transitions
observed in Fig.~\ref{fig:3} are lost.
In this case the vortex lattice can orient at
various angles with respect to the underlying
substrate.
In Fig.~\ref{fig:7}(a) we show the real space vortex positions
at $w/a = 2.8$, where the vortices form a triangular lattice
that is aligned at an angle $\theta=27.6^{\circ}$
with respect to the $y$-axis.
In Fig.~\ref{fig:7}(b) at $w/a = 3.272$,
$\theta = 24^{\circ}$, while in Fig.~\ref{fig:7}(c) at
$w/a = 3.53$,
$\theta = 2.9^{\circ}$, and in Fig.~\ref{fig:7}(d) at
$w/a = 3.75$,
$\theta = 13.2^{\circ}$.
In Ref.~\cite{79}, Guillamon {\it et al.} observed experimentally
that vortices on a q1D substrate retained triangular ordering
but that the vortex lattice was oriented at an angle $\theta$ ranging
from $\theta=0$ to $\theta=30^{\circ}$ with respect to the
substrate.
In several cases,
they found that the system locked to specific angles
close to $\theta=30^{\circ}$, $\theta=24^{\circ}$, and $\theta=0^{\circ}$.
We find a much larger variation in the orientation of the lattice
with respect to the
substrate as a function of filling than was observed
in the experiments, which may be due to differences the pinning strength
or the finite size of our simulations.
Our results show that for weak pinning,
the buckling transitions are lost and are replaced
with orientational transitions of the vortex
lattice with respect to the substrate.            
Another feature we observe when the pinning strength is increased
is that the vortex lattice becomes disordered or polycrystalline.
Guillamon {\it et al.} also observe that at higher fillings the vortex
lattice becomes disordered; however, in their system there are strong 
random vortex density fluctuations,
while for our thermally annealed samples the vortex density at
higher fields is generally uniform.     

\section{Dynamic Phases} 

\begin{figure}
\includegraphics[width=3.5in]{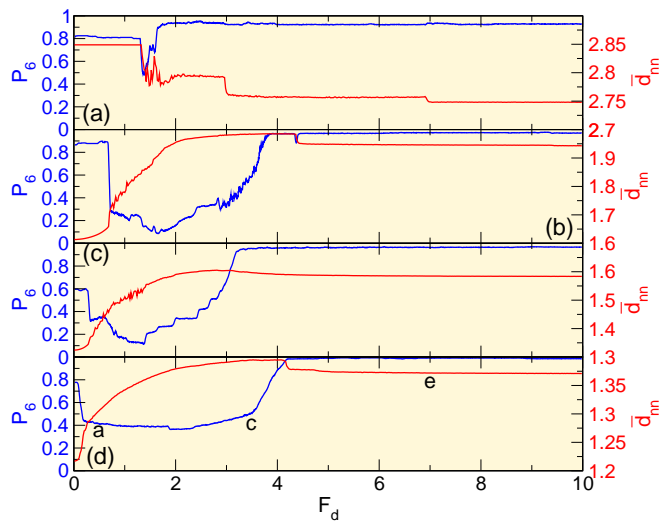}
\caption{
The fraction of six-fold coordinated vortices $P_{6}$ (dark blue curves)
and
the average nearest-neighbor distance $\bar d_{nn}$ (light red curves)
vs $F_{d}$
for a system with $F_{p} = 1.5$. 
(a) At $w/a  = 1.767$ the system depins from an $r_1$ state.
(b) At $w/a = 2.5$ the system dynamically orders into
a moving triangular lattice with $P_{6} = 0.97$. 
(c) $w/a = 3.061$.
(d) At $w/a = 3.535$, the onset of the dynamically ordered phase
coincides with a drop in $\bar d_{nn}$ near $F_{d} = 4.0$.
The labels a, c, and e correspond to the values of $F_{d}$
used for the images
in Fig.~\ref{fig:9}.     
}
\label{fig:8}
\end{figure}

In Fig.~\ref{fig:8} we plot simultaneously
$P_{6}$ and the average nearest neighbor spacing $\bar d_{nn}$
versus $F_{d}$
for a sample with $F_{p} = 1.5$ at varied $w/a$.
Here, we obtain $\bar d_{nn}$ by performing a Voronoi tesselation to
identify the $z_i$ nearest neighbors of particle $i$, and then
take
$\bar d_{nn}=(N \sum_i z_i)^{-1}\sum_{i=1}^{N}\sum_{j=1}^{z_i}r_{ij}$,
where $r_{ij}$ is the distance between particle $i$ and its $j$th nearest
neighbor.
For $w/a = 1.767$ in
Fig.~\ref{fig:8}(a), $P_{6}=0.83$ in the pinned $r_1$ state that occurs
for $0 < F_{d} < 1.4$.
There
is a dip in $P_{6}$ over the range  $1.4 < F_{d} < 1.7$,
corresponding to the plastic flow state in which
some of the vortices remain immobile while other vortices
hop in and out of the potential wells.
For $F_{d} > 1.7$, $P_{6}$ increases and reaches a saturated
value of $P_6=0.93$ when the vortices form a moving triangular lattice
containing
a small fraction of dislocations.
The value of $\bar d_{nn}$ drops at the depinning transition, and
several additional
drops in $\bar d_{nn}$ occur at higher drives.
In the $r_1$ pinned state,
each vortex has two close nearest neighbors that are in the same pinning
trough, and four more distant nearest neighbors that are in adjacent
pinning troughs.
Once the vortices depin and enter a moving state,
they adopt a more isotropic structure, causing
$\bar d_{nn}$ to drop as the distance to the
four more distant nearest neighbors decreases.
The additional drops in $\bar d_{nn}$ at higher
$F_{d}$ occur whenever the vortex lattice rearranges to become
still more isotropic.

\begin{figure}
\includegraphics[width=3.5in]{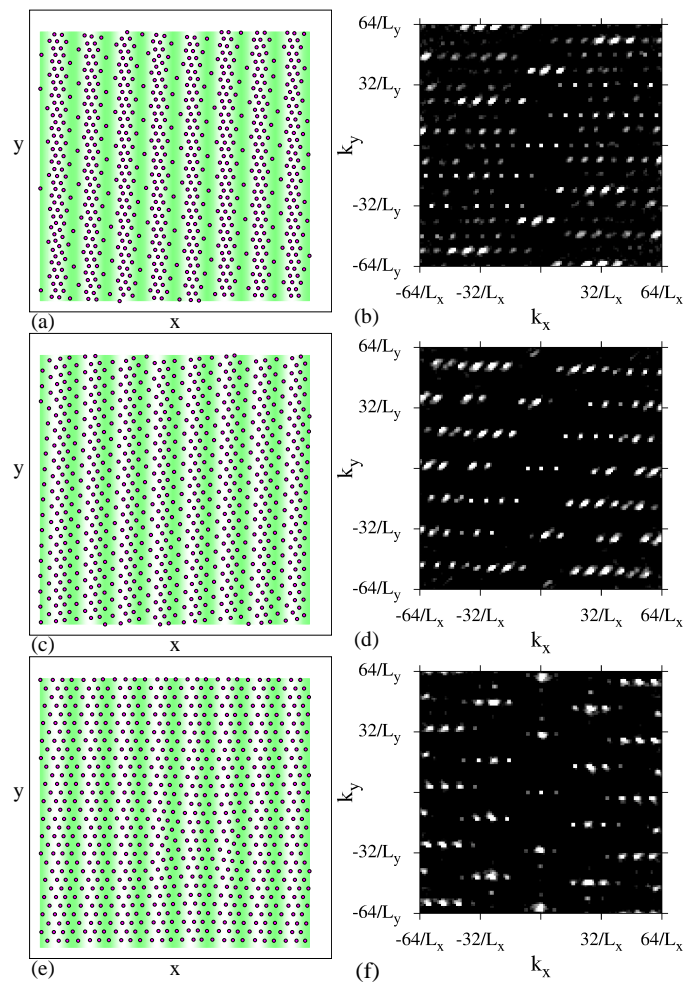}
\caption{
Real space images (left column), with the substrate minima indicated by
lighter regions and the vortex positions marked with circles, and $S(k)$
(right column)
for the dynamic system from Fig.~\ref{fig:8}(d) with $F_p=1.5$ and
$w/a=3.535$
at the values of $F_d$ labeled a, c, and e.
(a,b) At $F_{d} = 0.5$ the sample contains
pinned vortices coexisting with individual vortices
that hop from trough to trough.
(c,d) At $F_{d}  = 3.5$, all the vortices move together  to form 
a disordered lattice with a periodic density modulations.
(e,f) At $F_{d} = 7.0$ 
the system forms a moving floating triangular lattice.
}
\label{fig:9}
\end{figure}

\begin{figure*}
\includegraphics[width=4.5in]{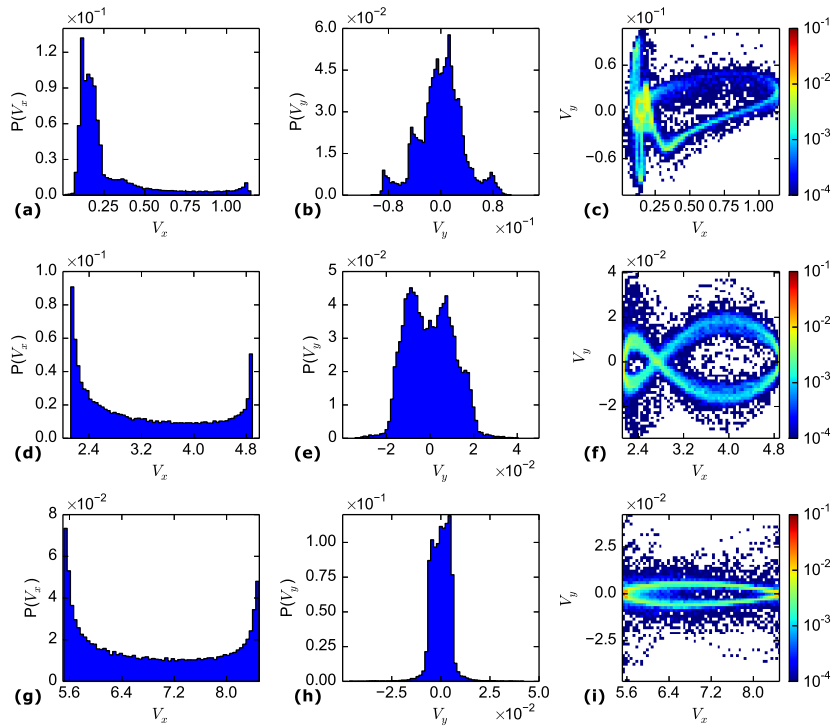}
\caption{
Vortex velocities for the system in Fig.~\ref{fig:9}
with $F_{p} = 1.5$ and $w/a = 3.535$. 
Left column: Histogram $P(V_x)$ of the instantaneous vortex velocities
in the driving direction $V_x$.
Center column: Histogram $P(V_y)$ of the instantaneous vortex velocities
in the transverse direction $V_y$.
Right column: Heightfield map of $V_y$ versus $V_x$.
(a,b,c) The plastic flow regime at $F_{d} = 0.5$.
(d,e,f) The moving modulated solid regime at $F_{d} = 3.5$.
(g,h,i) The moving floating solid regime at $F_{d} = 7.0$.
}
\label{fig:10}
\end{figure*}

In Fig.~\ref{fig:8}(b) we plot $P_6$ and $\bar d_{nn}$ versus $F_d$ for
the same sample at
$w/a = 2.5$ where an ordered zig-zag state
with $P_6=0.89$ appears at zero drive,
similar to that shown in Fig.~\ref{fig:1}(e).
The depinning threshold is
$F_{c} = 0.65$, much lower than the value of $F_c$ for the $w/a=1.767$ filling
in Fig.~\ref{fig:8}(a), 
and the
depinning transition is
marked by a drop in $P_{6}$ to $P_6=0.2$.
Over the range $0.65 < F_d \lesssim 2.0$, the vortices
are in
a dynamically disordered state,
while the saturation of $\bar d_{nn}$ above $F_d \approx 2.0$
indicates that a partially ordered state has formed.
The value of $P_{6}$ does 
not reach a maximum until $F_{d} = 3.9$,
where $P_{6} \approx 0.97$ and an ordered state appears.
For $1.5 < F_{d} < 3.9$, we observe a  moving density-modulated solid.  
The value of $F^{\rm tr}_c$, the drive at which the
sample reaches a moving triangular lattice
state, is higher for $w/a=2.5$ than for $w/a=1.767$, even though
the depinning threshold $F_c$ is smaller for the $w/a=2.5$ system.
At $w/a=2.5$, $\bar d_{nn}$ is initially small
and jumps up at the depinning transition,
unlike the decrease in $\bar d_{nn}$ at depinning found in
Fig~\ref{fig:8}(a).
Since the vortices in Fig.~\ref{fig:8}(b)
form a zig-zag $r_2$ structure
in the pinned state, each vortex has four close nearest neighbors in
the same pinning trough, and two more distant nearest neighbors
in an adjacent pinning trough.
This causes $\bar d_{nn}$ to be smaller in the pinned state than it was
for the $r_1$ structure in Fig.~\ref{fig:8}(a), and when the vortex
lattice becomes more isotropic in the moving state,
$\bar d_{nn}$ increases rather than decreasing as the two halves of each
zig-zag structure move further apart.
At $F_{d} = 4.1$ we observe a
drop in $\bar d_{nn}$ that coincides with a dip
in $P_{6}$.  This feature is associated with a transition
from a density-modulated lattice to a more uniform moving floating lattice.

In Fig.~\ref{fig:8}(c), we show $P_6$ and $\bar d_{nn}$ versus $F_d$
at $w/a = 3.061$ where there is an $r_3$ pinned state,
similar to that illustrated in Fig.~\ref{fig:2}(a).
Here the depinning threshold $F_{c} = 0.3$, and the system
transitions into a moving triangular lattice
at $F^{\rm tr}_{c} = 1.6$,
which is somewhat lower than the value of $F^{\rm tr}_c$
for $w/a = 2.5$ in Fig.~\ref{fig:8}(b).
The behavior of $\bar d_{nn}$ in Fig.~\ref{fig:8}(c) follows
a similar pattern as in Fig.~\ref{fig:8}(b),
with $\bar d_{nn}$ increasing with increasing $F_d$.
We plot the same quantities for $w/a=3.535$ in
Fig.~\ref{fig:8}(d),
where the depinning threshold
$F_c \approx 0.087$ and the system dynamically orders
for $F_{d} > 4.0$.
There is a small dip in $\bar d_{nn}$ at $F_{d} = 4.15$
along with a saturation in $P_6$ which is correlated with a
structural change to a dynamic floating lattice.

In order to characterize the nature of the dynamic
vortex structures in the moving states, in Fig.~\ref{fig:9}(a,b) 
we plot the real space images
and $S(k)$ for the system in
Fig.~\ref{fig:8}(d) at $w/a = 3.535$ and $F_{d} = 0.5$.
Here, individual vortices jump from one pinning well to the next while
a portion of the vortices
remain immobile in the substrate minima.
As shown in the plot of $S(k)$, the vortex configuration is fairly ordered
and takes the form of
a distorted non-triangular structure 
which causes $P_{6}$ to be low for this value of $F_{d}$.
For $F_{d} > F_{p}$, the vortices move together
so there is no plastic motion, and 
form a distorted lattice containing pronounced density modulations
as shown in Fig.~\ref{fig:9}(c,d) 
for $F_{d} = 3.5$.
For $F_{d} > 4.0$ we find a transition from  the density modulated lattice 
to a moving homogeneous floating  triangular lattice which 
coincides with the drop in $\bar d_{nn}$ in
Fig.~\ref{fig:8}(d) and the maximum in $P_{6}$.
Fig.~\ref{fig:9}(e) 
shows the floating lattice 
at $F_{d} = 7.0$, where as indicated in
Fig.~\ref{fig:9}(f) $S(k)$ contains sixfold peaks that are indicative
of triangular ordering.
The smectic ordering induced by the substrate is substantially weaker
or almost absent at this drive, as shown by 
the weakness of the spots in $S(k)$ at $k_{y} = 0$,
indicating that the system has formed a floating solid. 
We find similar types of transitions in the dynamics at other fillings as well. 

\begin{figure}
\includegraphics[width=3.5in]{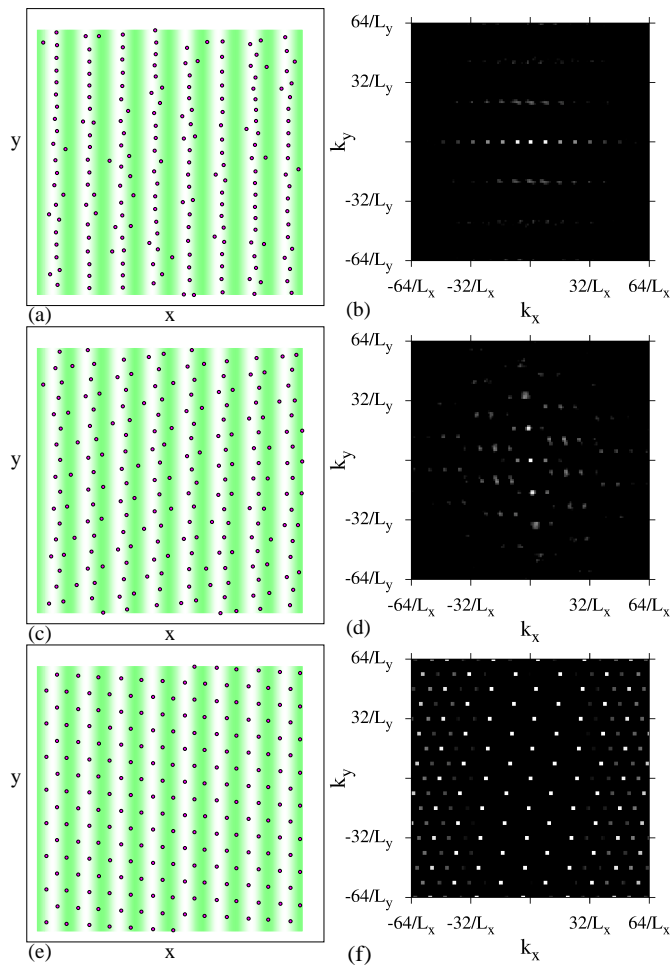}
\caption{
Real space images (left column), with the substrate minima indicated by
lighter regions and the vortex positions marked with circles, and  
$S(k)$ (right column) for a
system with $F_{p} = 0.5$ and $w/a = 1.767$.
(a,b) The plastic flow phase at $F_{d} = 0.3$ where there
is individual vortex
hopping from well to well. The $S(k)$ peaks indicate a smectic phase
with periodic ordering along the $x-$direction.
(c,d) At $F_{d} = 0.6$, all the vortices
are flowing and form chains that are aligned in the $x$-direction.
$S(k)$ shows that a new smectic order has appeared
with periodic ordering along the $y$-direction. 
(e,f) At $F_{d} = 3.6$ there is a moving floating triangular crystal.   
}
\label{fig:11}
\end{figure}

\begin{figure*}
\includegraphics[width=4.5in]{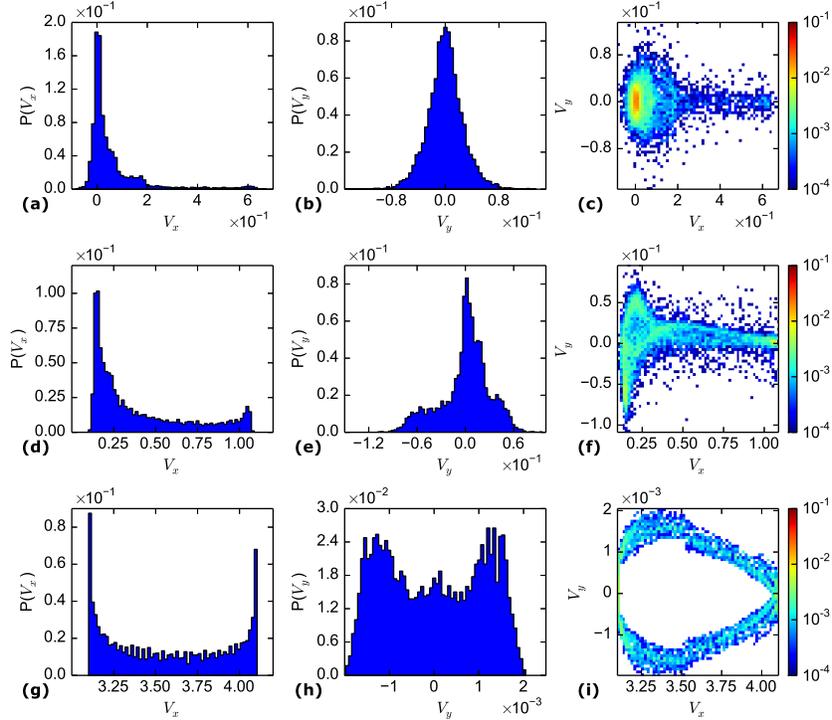}
\caption{
Vortex velocities for the system in Fig.~\ref{fig:11} with
$F_p=0.5$ and $w/a=1.767$.
Left column: $P(V_x)$.  Center column: $P(V_y)$.  Right column:
Heightfield map of $V_y$ versus $V_x$.
(a,b,c) The plastic flow phase at $F_{d} =0.3$
where there is a large peak in $P(V_x)$ at $V_{x} = 0$
due to the pinned vortices.
(d,e,f) The moving smectic phase from
Fig.~\ref{fig:11}(c,d) at $F_{d} = 0.6$.
(g,h,i) The moving triangular solid phase at $F_{d} = 3.6$.
}
\label{fig:12}
\end{figure*}

We can also characterize the different dynamic states in
Figs.~\ref{fig:8} and \ref{fig:9} by examining histograms of the
vortex velocities.
In Fig.~\ref{fig:10}(a) we plot the distribution $P(V_x)$ of
$V_x$ in the driving direction
at $F_{d} = 0.5$ for the system in 
Fig.~\ref{fig:9}(a,b) with $F_{p} = 1.5$ and $w/a = 3.535$. 
Figure~\ref{fig:10}(b) shows
the transverse velocities $P(V_{y})$, while
in Fig.~\ref{fig:10}(c) we plot $V_{y}$ versus $V_{x}$ as a heightfield map.  
At this drive, the motion is plastic and occurs by individual vortex hopping, 
so there is a sharp peak in $P(V_{x})$ at
$V_x=0.15$ which reflects 
the fact that most of the vortices
are slowly moving within an individual pinning trough.
When a single vortex jumps into an adjacent pinning trough,
it creates a pulse of motion through the trapped vortices
that triggers the jump of another single vortex to the next pinning
trough, where the process repeats.
This depinning cycle creates
two peaks in $P(V_x)$.
The peak at low $V_x$
corresponds to the motion of a velocity pulse through the dense
assembly of vortices at the bottom of the pinning trough,
while the peak at high $V_x$ is produced by individual vortices
escaping over the potential maximum.
This peak falls near $V_x=1.1$, which is larger than $F_d$, reflecting
the fact that after an individual
vortex passes the crest of the substrate maximum,
the substrate contributes an additional force term in the driving
direction as the vortex moves toward the next substrate minimum.
In this case the maximum force exerted by the pinning site
is $F_p=1.5$ while the driving force is $F_d=0.5$,
so that the maximum possible instantaneous vortex velocity
would be $V_x=2.0$; however, vortex-vortex interactions prevent individual
vortices from moving this rapidly.
In Fig.~\ref{fig:10}(b), $P(V_y)$
is centered at $V_y=0$
since there is no driving force in the transverse direction;
however, we observe
some asymmetry in $P(V_{y})$ as well as
peaks at finite $V_{y}$ 
due to the fact that the vortex lattice segments inside the pinning
troughs are oriented at an angle with respect to the
substrate symmetry direction, as shown
in Fig.~\ref{fig:9}(a).
This asymmetry also appears in the $V_{y}$ versus $V_{x}$ plot
in Fig.~\ref{fig:10}(c), which has two prominent features.
The first is a wide band of $V_y$ values 
at low $V_{x}$ that are
associated with soliton-like
pulses moving through the dense regions of the vortex clusters,
which push vortices
in both the positive and negative $y$-direction.
The second is the loop shape at larger $V_x$ values which corresponds to
motion in which the vortices are accelerated or decelerated as
they pass over the
substrate maxima and minima.

In Fig.~\ref{fig:10}(d,e,f) we show instantaneous velocity plots
for the system in 
Fig.~\ref{fig:9}(c,d) with $F_{d} = 3.5$ where the vortices are
moving elastically in the density-modulated solid phase.
Here 
$P(V_{x})$ in Fig.~\ref{fig:10}(d) has peaks 
at $V_x=2.01$ and $V_x=4.75$ which are smoothly connected by finite
$P(V_x)$ values.
The shape of this histogram shows
the velocity imposed by the driving force of $F_{d} = 3.5$ when
the substrate forces alternately act with or against the driving force.
When the substrate force is against the drive 
the velocity is $V_x=F_{d} - F_{p} = 2.0$,
while when the substrate and driving forces are in the same
direction, 
$V_x=F_{d} + F_{p} = 5.0$, close to the observed
values of the peaks in $P(V_x)$. 
In Fig.~\ref{fig:10}(e),
$P(V_{y})$ has two peaks close to
$V_y=0.1$ and $V_y=-0.1$, indicating that there is 
an oscillatory motion in the $y$-direction. 
This effect can be seen more clearly in
the $V_{y}$ versus $V_{x}$ plot in Fig.~\ref{fig:10}(f)
which has two symmetric lobes.
Since the vortices
are in a density-modulated lattice,
shearing in the $y$-direction occurs between adjacent density modulations, 
with one density modulation 
moving in the positive $y$-direction
while the other moves in the negative $y$-direction.

The velocity plots
for the system in Fig.~\ref{fig:9}(e,f) at $F_{d} = 7.0$ appear in
Fig.~\ref{fig:10}(g,h,i).
In Fig.~\ref{fig:10}(g),
$P(V_{x})$  
has a two-peak feature similar to that in
Fig.~\ref{fig:10}(d),
but with peak values at $V_x=5.52$ and $V_x=8.4$.
Figure ~\ref{fig:10}(h) shows that
$P(V_{y})$ has a single peak centered at $V_{y} = 0$,
while in Fig.~\ref{fig:10}(i), there is a
single lobe in the $V_{y}$ versus $V_{x}$ plot.
Here the vortices
have formed a floating triangular solid, and 
their motion is close to one-dimensional along the driving direction.
As $F_{d}$ is further increased, the 
width of the lobe feature in the $V_y$ direction
gradually decreases.
We find similar histograms for the other fillings in the strong pinning limit
for the plastic flow, moving modulated solid, and moving floating solid
regimes.      

\subsection{Smectic to Smectic Transitions}

\begin{figure}
\includegraphics[width=3.5in]{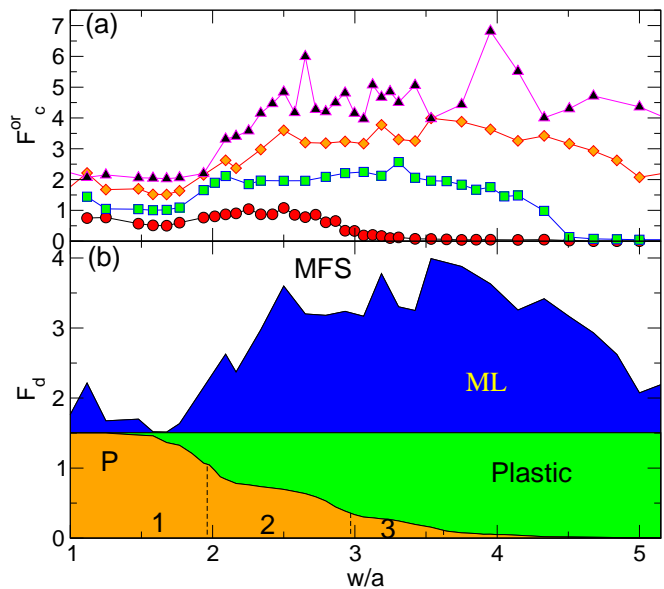}
\caption{
(a) $F^{\rm or}_c$, the drive at which the system transitions from a
density modulated moving crystal to an ordered moving floating solid,
vs $w/a$ at
$F_p=$ 0.5 (red circles),
1.0 (green squares),
1.5 (orange diamonds),
and 2.0 (black triangles).
(b) Dynamic phase diagram as a function of $F_d$ and $w/a$ for
a system with $F_p=1.5$.  P: pinned phase.  Plastic: plastic flow regime.
ML: moving modulated lattice state.  MFS: moving flowing solid state.
Dashed lines are guides to the eye that indicate the transition from
$r_1$ to $r_2$, $r_2$ to $r_3$, and $r_3$ to a disordered pinned state.
}
\label{fig:13}
\end{figure}

For $F_{p} < 1.0$, we find that 
a dynamically induced smectic to smectic transition can occur. 
In Fig.~\ref{fig:11}(a,b) we show the real space and
$S(k)$ images for a system with $F_{p} = 0.5$ and
$w/a = 1.767$ in the plastic flow regime
where there is a combination of vortices that are trapped
in the pinning troughs and a smaller amount of vortices that hop by
jumping from one trough to the next and then triggering a jump of
another vortex from one trough to the next.
Here, $S(k)$ indicates that
the overall system has smectic ordering
due to the chain-like structure of the vortices within the pinning troughs.
In Fig.~\ref{fig:12}(a) we plot $P(V_{x})$ at $F_{d} = 0.3$,
where there is a peak at $V_x=0$ due to the
pinned vortices
along with a small bump at $V_{x} = 0.6$ due to the vortex hopping.
Figure~\ref{fig:12}(b) shows that $P(V_{y})$ has a maximum
at $V_{y} = 0$, while in Fig.~\ref{fig:12}(c),
the $V_{y}$ versus $V_{x}$ plot is asymmetric in $V_{y}$,
with a peak
at $V_{x} = V_{y} = 0.0$ and a second peak at
higher $V_{x}$ produced by the moving vortices.  

In Fig.~\ref{fig:11}(c) we show the real space
vortex configuration at $F_{d} = 0.6$, which is higher than the maximum pinning
force of $F_{p} =0.5$.
All the vortices are in motion, but instead of retaining their alignment
along the $y$-direction induced by the substrate,
they form a chain-like structure
aligned in the $x$-direction with a slight tilt in the positive
$y$-direction.
This alignment in the drive direction
is more clearly seen in the corresponding $S(k)$ in Fig.~\ref{fig:11}(d),
where the peaks fall 
along $k_{x}= 0$, indicating a smectic phase
with ordering along the $y$-direction. There are some very weak
peaks on the $k_{y} = 0$ axis due to the substrate,
but overall the vortex structure is a smectic
state rotated $90^{\circ}$ from the $y$-axis. 
The peaks do not fall exactly at $k_{x} = 0$ but are at a slight angle,
due to the channels 
in Fig.~\ref{fig:11}(c) being slightly tilted in
the positive $y$-direction.
In Fig.~\ref{fig:12}(d),
$P(V_{x})$ for this case shows a peak at
$V_{x} = 0.19$, while there is an absence of
weight in $P(V_x)$ at $V_{x} = 0.0$, 
indicating that the vortices are always in motion.  
Figure~\ref{fig:12}(e) shows that
$P(V_{y})$ peaks at $V_{y} = 0.0$ and has an overall asymmetry,
which also appears in the $V_y$ versus $V_x$ plot in
Fig.~\ref{fig:12}(f).
The vortex channeling occurs when the vortices form
effective pairs aligned along the $x$-direction.
In each pair, one vortex is slowed by the backward-sloping side of the
potential trough, while the other vortex is sped up by the
forward-sloping side of the potential.  The faster vortex pushes the
slower vortex, giving the pair an increased net motion along the
$x$ direction.
This pairing effect is visible in the real space image in Fig.~\ref{fig:11}(c).

As $F_{d}$ further increases, there is a transition to a flowing solid phase
as shown
in Fig.~\ref{fig:11}(e,f) for $F_{d} = 3.6$,
where $S(k)$ has sixfold ordering.     
The corresponding $P(V_{x})$
in Fig.~\ref{fig:12}(g)
has two peaks, while
$P(V_{y})$ in Fig.~\ref{fig:12}(h)
has a symmetrical distribution with three peaks indicating that there is
an oscillation in the vortex orbits in the $y$-direction.
The plot of $V_y$ versus $V_x$ in
Fig.~\ref{fig:12}(h) contains
a single lobe similar to that found
for the moving floating solid in Fig.~\ref{fig:10}(i).
As $F_{d}$ is increased still further,
the width of this lobe in the $V_y$ direction decreases.
The smectic-to-smectic transition is 
limited to
the range $1.5 < w/a < 2$, in which two vortices can fit between
adjacent potential maxima in the dynamically moving regime.

\section{Dynamical Phase Diagram} 

For $F_{p} > 0.25$, the drive $F^{\rm or}_{c}$ at which
the system transitions from a density modulated moving crystal to 
an ordered moving floating solid shows considerable variation with $w/a$,
particularly for the
larger values of $F_{p}$.
In Fig.~\ref{fig:13}(a) we plot $F^{\rm or}_{c}$, determined from
the location of a feature in $P_6$, 
versus $w/a$ for $F_{p} = 0.5$, 1.0, 1.5, and $2.0$.
For $F_{p} = 0.5$, $F^{\rm or}_c$ has a local maximum 
near  $w/a = 2.5$, and then
drops for $w/a > 3.0$.
In the pinned phase
for $w/a >3.0$, the system forms a polycrystalline state,
and in the moving state
the grains realign to form a moving crystalline state. 
For $F_{p} = 1.0$, 1.5, and $2.0$,
when $w/a < 1.75$ the system depins from a single
chain of vortices and can partially form a moving crystal state.
When $w/a$ is large enough that a pinned zig-zag state
forms, the moving density-modulated state can persist
up to much higher drives.
For $F_{p} = 1.0$ the system forms a pinned polycrystalline
state for $w/a > 4.5$ which coincides
with the drop in $F^{\rm or}_{c}$ for $w/a > 4.5$.
For $F_{p} = 1.5$ and $2.0$,  there are local peaks in 
$F^{\rm or}_{c}$ that correlate with 
the moving buckled phases
which occur when groups of vortices can
fit between adjacent pinning maxima  as the vortices move.
This effect is most pronounced for $F_{p} = 2.0$.

In Fig.~\ref{fig:13}(b)
we plot a dynamic phase diagram as a function of $F_{d}$
and $w/a$ for a system with $F_{p} = 1.5$.
Above depinning in the regime where $F_{d} < F_{p}$, the
system is in a plastic flow state
in which there is a coexistence of moving vortices and immobile vortices.
In this regime the structure factor
generally shows disordered features.
For $F_{d} > F_{p}$, all the vortices are moving and the system
is either in a modulated
lattice (ML) state or a moving flowing solid (MFS) state.
We find similar dynamical phase diagrams for other
values of $F_{p} > 1.0$,
while for the weaker substrates,
the size of the plastic flow region is reduced and the ML phase is replaced
with a smectic moving state similar to that shown in
Fig.~\ref{fig:11}(c,d).     

\section{Discussion}

The dynamic phases we observe have certain
similarities to the dynamic states observed for vortices moving over random
pinning arrays in that there can be pinned, plastic, and
dynamically ordered phases
as a function of external drive \cite{80,81,82,83,84,85}.
There are some differences, including the fact
that the moving modulated lattice we observe does not form a
smectic state that is aligned in the drive direction,
as found for vortices moving over random
pinning arrays \cite{82,83,84,85}.
Future studies might consider combinations of random 
disorder with periodic disorder,
which would introduce a competition in the moving phase
between the smectic ordering
imposed by the substrate and the smectic ordering induced by the drive.  
Additionally, for random pinning arrays simulations
indicate that for increasing vortex density,
the drive at which the transition to the ordered state occurs decreases
due to the increase in the strength of the
vortex-vortex interactions \cite{85}.
For the q1D substrate,
the location of the ordering transition shows strong fluctuations
due to the ability of the moving lattice to become
dynamically commensurate with
the periodicity of the substrate. 

\section{Summary} 

We examine the statics and dynamics of vortices interacting with 
a periodic quasi-one-dimensional substrate
in the limit where the vortex lattice
spacing is smaller than the spacing of the periodic lattice.
For weak substrate strengths, we find that the vortices
retain hexagonal ordering but 
exhibit numerous rotations with respect to the substrate,
similar to recent experimental observations.
For stronger substrates there are
a series of buckling transitions where the vortices can form
anisotropic 1D chains, zig-zag patterns, and higher order numbers of chains
within each substrate minimum. 
At some fillings the overall lattice has long range order and becomes
partially distorted at the transitions between these states.
For higher fillings the buckling
transitions are lost and the system forms a polycrystalline state.
We also find that the depinning
shows a series of step like features
when the system transitions from a state with $n$ chains
to a state with $n+1$ chains in each substrate minimum,
and that for weaker pinning there are some cases where there is a peak 
in the depinning force as a function of filling.
For weak substrates, under an applied drive 
the vortices depin elastically and 
retain their triangular ordering,
while for strong substrates the buckled states transition
to a partially disordered flowing state
followed by various other transitions into moving 
modulated crystal or homogeneous floating moving crystal states.
Our results should also be applicable to other systems of
particles with repulsive interactions
in the presence of a periodic quasi-one dimensional substrate,
such as electron crystals, colloids, and ions in optical traps.                

\acknowledgments
This work was carried out under the auspices of the 
NNSA of the 
U.S. DoE
at 
LANL
under Contract No.
DE-AC52-06NA25396.
The work of DM was supported in part by the U.S. Department of Energy,
Office of Science, Office of Workforce Development for Teachers and
Scientists (WDTS) under the Visiting Faculty Program (VFP).

\end{document}